**Estimating Heterogeneity in Travel Mode Choice Shifts with Causal Forests**

**Rishabh S. Chauhan**
Industry Assistant Professor
Center for Urban Science and Progress, Tandon School of Engineering
New York University, Brooklyn, NY 11201
Email: rsc9107@nyu.edu

**Mahdi Ghadimi***
Graduate Research Assistant
Center for Urban Science and Progress, Tandon School of Engineering
New York University, Brooklyn, NY 11201
Email: mg8786@nyu.edu

**Lishun Liu Gaara**
Graduate Research Assistant
Center for Urban Science and Progress, Tandon School of Engineering
New York University, Brooklyn, NY 11201
Email: ll5548@nyu.edu

Total Number of Pages: 20

*Corresponding Author

## ABSTRACT

**Objectives:** While causal analysis of travel behavior is an emerging field, estimating heterogeneity in mode choice through causal modeling remains unexplored. This study demonstrates the application of a novel causal method, causal forest, to quantify the heterogeneity in travel mode choice shifts caused by the COVID-19 pandemic.

**Methods:** We applied causal forests, a non-parametric causal machine learning method, to 802,935 trip records from the 2017 and 2022 waves of the National Household Travel Survey. The 2017 wave serves as the pre-pandemic control group, while the 2022 wave represents the treatment condition. Within the potential outcomes framework, we estimate average treatment effects (ATE), heterogeneous treatment effects (HTE), and conditional average treatment effects (CATE) across diverse socio-demographic groups and trip characteristics.

**Findings:** Our results reveal an estimated ATE of a 1.86 percentage point (pp) increase in car-mode share, contrasted with decreases of 0.38 pp and 1.57 pp in public transit and walking, respectively. The largest increases in car use appeared for short-distance trips (one mile or less), households with annual incomes exceeding $200,000, and female travelers.

**Novelty:** This is one of the first applications of causal forests to travel mode choice, and the first to use causal machine learning to estimate the pandemic's causal effect on mode choice analysis.

**Practical Applications:** This study discusses methodological advantages, inherent assumptions, and limitations of causal forests within the context of transportation planning. This methodology is applied to COVID-19 travel data to illustrate how causal heterogeneity analysis can offer a deeper understanding of changes in mode choice. These insights are valuable for planners and policymakers in making policies related to mode shifts under an intervention.

## INTRODUCTION

Travel behavior analysis is crucial for effective transportation planning. It has traditionally been conducted using discrete choice modeling and random utility maximization (McFadden 2001; Derrible 2025). Such a statistically driven framework allows researchers to quantify associations between demographics, built environments, and mode choices. As computational capabilities expanded, machine learning and deep learning algorithms, particularly artificial neural networks and gradient boosting machines, emerged as promising alternatives. These advanced algorithms drastically improve predictive accuracy by capturing complex, non-linear relationships from travel data (Karlaftis and Vlahogianni 2011). However, almost all statistical, machine learning, and deep learning methods suffer from a fundamental conceptual roadblock: causality (Chauhan, Riis, et al. 2024; Pearl and Mackenzie 2018). These methods are built to study correlations (or associations), not causation, and thus, they cannot explain *why* something happened, differentiate between cause and effect, or estimate the effect of an intervention (Pearl and Mackenzie 2018).

In fields such as technology and medical science, randomized controlled trials, frequently implemented as A/B testing, represent the gold standard for extracting unconfounded causal effects (Imbens and Rubin 2015; Kohavi et al. 2009). However, conducting large-scale randomized trials to analyze travel behavior would likely be infeasible and unethical. A preferred approach would estimate causal effects directly from observational data. As recently demonstrated by Chauhan et al. (2026) causal discovery and inference methods provide valuable opportunities to examine mode choice within a causal framework. This approach allows tracing the decision-making process through a structural causal graph and estimating quantitative causal effects from travel survey data.

Still, a global causal effect relies on broad generalization of the population; however, travel behavior is notoriously heterogeneous. Different socioeconomic demographic groups have different Value of Time (VOT), affordability, and travel preferences (Hossan et al. 2016). Incorporating heterogeneity in causal mode choice analysis allows for different estimates across various population groups, reflecting the reality that different demographic groups make choices in unique ways. To fill this critical methodological gap, this study employs a recently developed nonparametric causal machine learning method, known as causal forests, to study mode choice. Causal forests are grounded in causality and are recognized for their ability to estimate heterogeneous treatment effects (Athey and Wager 2019).

The COVID-19 pandemic serves as a significant case study for this analysis due to its unparalleled impact on global mobility, drastically altering travel behavior worldwide (De Vos 2020; Salon, Conway, Capasso Da Silva, et al. 2021). This disruption provides a unique opportunity to demonstrate the effectiveness of causal forests in estimating heterogeneous causal effects.

By analyzing over 800,000 trip records from the 2017 and 2022 waves of the National Household Travel Survey (NHTS), we utilize Causal Forests to isolate the true treatment effect of the pandemic on car, public transit, and walking mode choices. We focus on estimating heterogeneous treatment effects across key sociodemographic and trip characteristics to show which groups changed their travel behavior and by how much.

## LITERATURE REVIEW

Although causality may be an intuitive concept, its formal development as a computational framework is relatively recent. The earliest application of potential outcomes can be traced to Neyman's study of a hypothetical agricultural experiment (Neyman 1923; Rubin 2005). Rubin (Rubin 1974) subsequently extended Neyman's notations to formalize the potential outcomes framework, enabling the estimation of causal effects by comparing outcomes under treatment versus control. The foundation for using propensity scores to estimate causal effects in observational studies was later laid by Rosenbaum and Rubin (Rosenbaum and Rubin 1983). Furthermore, Pearl (Pearl 2000) introduced Structural Causal Models (SCM), establishing a mathematical framework for analyzing causes and counterfactuals. In recent decades, the causal effects of interventions have been evaluated using observational data across fields such as medicine, epidemiology, biostatistics, economics, and other social sciences (Dahabreh and Bibbins-Domingo 2024).

Causal analysis has seen very limited application in the context of travel mode choice. Initial efforts involved using Bayesian Networks for mode choice analysis (Xie and Waller 2010; Ma et al. 2017). Later, Monteiro (Monteiro 2020) applied a causal discovery algorithm to construct a mode choice causal graph. To estimate both causal graphs and causal effects, Chauhan et al. (2024) introduced a framework integrating causal discovery algorithms with Structural Equation Models (SEM). More recently, researchers have combined causal discovery with causal inference techniques to investigate mode choice behavior (Chauhan et al. 2026, 2025). Additionally, quasi-experimental approaches, including propensity score matching and difference-in-differences, have been utilized to evaluate the effects of specific interventions, like the addition of a new metro line or stations, on travel mode choice (Chiu 2022; Dai et al. 2020).

To evaluate heterogeneity in causal effects, Athey and Imbens (Athey and Imbens 2016) developed the causal trees methodology. Wager and Athey (Wager and Athey 2018) extended this idea to develop causal forests based on the random forest framework. Causal forests are nonparametric and can estimate heterogeneous treatment effects (HTE). Despite these advantages, causal forests have rarely been used in travel mode analysis. As per our knowledge, Blättler et al. (2024) is the only study to do so. Blättler et al. (2024) examined how a fare-free public transit policy for overnight visitors affected travel mode choice to a tourist destination in Switzerland. However, their research was restricted in scope, focusing exclusively on tourism-related travel to a single location. Furthermore, while their work analyzed conditional average treatment effects (CATE) among different guest categories, it yielded only positive outcomes, and they offered limited insight into how population heterogeneity manifests across distinct trip types and socio-demographic variables.

The COVID-19 pandemic was a major disruption to travel, driven by changes such as remote work, online shopping, risk perception, and travel restrictions. The effects of the pandemic on travel behavior are extensively studied (Drummond and Hasnine 2022; Conway et al. 2020; Shamshiripour et al. 2020; Salon et al. 2022; Mohammadi et al. 2022). Longitudinal surveys were conducted to track the evolution of these behavioral shifts (Chauhan, Bhagat-Conway, et al. 2024; Chauhan, Conway, et al. 2021; Salon et al. 2020). Although these studies provide deep insight into travel mode choice shifts, the existing literature relies on correlation-based methodologies (Salon, Conway, Capasso Da Silva, et al. 2021; Javadinasr et al. 2022; Magassy et al. 2023), leaving a critical gap in causal understanding.

Our study demonstrates the application of causal forests to analyze the heterogeneous shifts in travel mode choice. We apply this methodology to examine the shifts in travel mode choices caused by the COVID-19 pandemic. Our research fills the gap in literature in two ways: (i) It is one of the first applications of causal forests within the domain of travel mode choice analysis. (ii) To the best of our knowledge, this work is the first attempt to implement a causal inference framework, specifically causal machine learning, to evaluate the causal impact of the pandemic on mode choices. (iii) This article offers guidance for future research on the effective application, interpretation, strengths, and limitations of this methodology for mode choice analysis.

## METHODOLOGY

### Survey Data and Sample Selection

This study utilizes the 2017 and 2022 National Household Travel Survey (NHTS) datasets (Federal Highway Administration 2022). The 2017 wave establishes the pre-pandemic control baseline, while the 2022 wave captures the pandemic-era treatment condition. The 2022 wave was conducted between January 2022 and January 2023, which is later in the pandemic when travel began to rebound (International Trade Administration, n.d.). Thus, the NHTS 2022 wave was deemed an appropriate proxy for more stable travel behavior later in the pandemic.

During initial data processing, we removed any trip records containing indeterminate responses (e.g., "I don't know" or "appropriate skip"). We also strictly limited the analytical scope to adult decision-makers, thereby excluding all respondents under 18. To distinguish the temporal cohorts, we constructed a binary treatment indicator $W_i$. Trips originating from the 2017 baseline received a value of $W_i = 0$. Trips from the 2022 wave received $W_i = 1$. The final harmonized dataset encompasses 802,935 independent trip observations.

The analytical outcome space centers on three binary dependent variables. Each represents a primary travel mode:

I. *Car_x*: Drove a private vehicle (car, pickup truck, SUV, or van).
II. *Public_x*: Used shared public transit (bus, subway, commuter rail, streetcar, or Amtrak).
III. *Walk_x*: Walked as the primary mode of travel.

**Variables and Covariates**

To truly isolate the pandemic's temporal shock, the estimation must be conditioned on a comprehensive matrix of exogenous covariates. As detailed in Table 1, the feature space spans geographic, personal, household, and trip-level dimensions. Passing these specific vectors into the causal algorithm controls for underlying demographic drift over the five-year survey gap. This ensures that any estimated modal shift is attributable to the pandemic era rather than to sampling variance.

**TABLE 1 Descriptive Statistics and Covariate Definitions**

| Group | Variable | Description | Category | Interpretation | % |
|---|---|---|---|---|---|
| Place characteristic | Region | Census region classification for home address | 1 | Northeast | 15.95 |
| | | | 2 | Midwest | 16.00 |
| | | | 3 | South | 43.05 |
| | | | 4 | West | 25.00 |
| | Urban area | Household urban area classification, based on 2020 TIGER/Line Shapefile | 0 | Outside of urban area or urban cluster | 21.68 |
| | | | 1 | In an urban area or urban cluster | 78.32 |
| Personal characteristic | Sex | Male | 0 | No | 53.80 |
| | | | 1 | Yes | 46.20 |
| | Age | Traveler's age (years) | 1 | 18-29 | 10.15 |
| | | | 2 | 30-45 | 22.60 |
| | | | 3 | 46-65 | 40.57 |
| | | | 4 | 65+ | 26.68 |
| | Education | Educational Attainment | 1 | Less than a high school graduate | 2.83 |
| | | | 2 | High school graduate or GED | 17.02 |
| | | | 3 | Some college or associates degree | 29.29 |
| | | | 4 | Bachelor's degree | 26.67 |
| | | | 5 | Graduate degree or professional degree | 24.19 |
| Household characteristics | Household income | Household yearly income (USD) | 1 | $24,999 or less | 13.22 |
| | | | 2 | $25,000 - $49,999 | 19.49 |
| | | | 3 | $50,000 - $99,999 | 33.13 |
| | | | 4 | $100,000 - $199,999 | 26.35 |
| | | | 5 | $200,000 or more | 7.82 |
| | Household size | Count of household members | 1 | 1 | 18.97 |
| | | | 2 | 2 | 46.10 |
| | | | 3 | 3 or more | 34.93 |
| | Vehicle ownership | Count of Household vehicles | 0 | 0 | 2.33 |
| | | | 1 | 1 | 23.61 |

| | | | 2 | 2 or more | 74.07 |
|---|---|---|---|---|---|
| Trip characteristics | Trip distance | Trip distance (miles) | 1 | ≤ 1 | 19.19 |
| | | | 2 | > 1 and ≤ 5 | 40.70 |
| | | | 3 | > 5 and ≤ 20 | 30.29 |
| | | | 4 | > 20 and ≤ 50 | 7.34 |
| | | | 5 | > 50 | 2.49 |
| | Trip purpose | Purpose of trip | 0 | Not home-based | 34.08 |
| | | | 1 | Home-based | 65.92 |
| | Weekend trip | Trip of weekend | 0 | No | 76.64 |
| | | | 1 | Yes | 23.36 |
| | Year | Year of data collection | 0 | 2017 | 96.76 |
| | | | 1 | 2022 | 3.24 |
| Mode choice | Public transit | Used Public Transit (Public or commuter bus, Subway or elevated rail, Commuter rail, Street car or trolley car, or Amtrak) | 0 | No | 98.71 |
| | | | 1 | Yes | 1.29 |
| | Car | Drove a private vehicle (Car, Pickup truck, SUV, or Van) | 0 | No | 10.16 |
| | | | 1 | Yes | 89.84 |
| | Walking | Walk | 0 | No | 91.13 |
| | | | 1 | Yes | 8.87 |
| Total number of observations | | | | 802,935 | |

**Potential Outcomes Framework**
This analysis utilizes the Rubin Causal Model (Rubin 1974) to formally identify causal effects from observational cross-sections. Let $i = 1, \dots, N$ index the individual trip observations. Let $W_i \in \{0,1\}$ denote the binary treatment assignment. For each unit $i$, the framework posits two distinct potential outcomes. $Y_i(1)$ denotes the mode choice decision if exposed to the pandemic environment, and $Y_i(0)$ denotes the mode choice decision if the unit $i$ existed in the pre-pandemic baseline.

The fundamental causal effect of the temporal shock on an individual unit is defined as the arithmetic difference between these potential outcomes in Equation 1:

$$\tau_i = Y_i(1) - Y_i(0) \tag{1}$$

As Holland (Holland 1986) formalized, observing both potential outcomes for a single unit simultaneously is impossible. Therefore, the observed outcome $Y_i$ functions as a combination of the potential outcomes mapped to the realized treatment assignment, as shown in Equation 2:

$$Y_i = W_i Y_i(1) + (1 - W_i) Y_i(0) \tag{2}$$

Because individual-level effects ($\tau_i$) contain unobservable counterfactuals, estimation requires several assumptions. First, Unconfoundedness assumes that, conditional on the pre-treatment covariate vector $X_i$, the treatment assignment is statistically independent of the potential outcomes as formalized in Equation 3:

$$\{Y_i(0), Y_i(1)\} \perp\!\!\!\perp W_i \mid X_i \tag{3}$$

Second, the Overlap (Positivity) assumption mandates that every covariate profile $x$ maintains a non-zero probability of assignment to either condition (Equation 4):

$$0 < P(W_i = 1 \mid X_i = x) < 1 \quad \forall x \in X \tag{4}$$

**Average Treatment Effect and Selection Bias**

Observational survey data inherently contain selection bias due to changing demographic compositions across sampling waves. As demonstrated by Holland (1986), a naive difference in observed means actually yields a composite of the true causal effect and selection bias (Equation 5):

$$E[Y \mid W = 1] - E[Y \mid W = 0] = \underbrace{E[Y(1) - Y(0) \mid W = 1]}_{ATT} + \underbrace{(E[Y(0) \mid W = 1] - E[Y(0) \mid W = 0])}_{Selection\ Bias} \quad (5)$$

In this equation, $E[Y(1) - Y(0) \mid W = 1]$ is defined as the Average Treatment Effect on the Treated (ATT). Rigorous causal frameworks seek to isolate this ATT by removing the selection bias under appropriate identification assumptions.

The Average Treatment Effect (ATE) represents the expected mean difference between the potential outcome under treatment and control outcomes across the population, as presented in Equation 6 (Rubin 1974):

$$\tau_{ATE} = E[Y_i(1) - Y_i(0)] \quad (6)$$

The causal forest first estimates Conditional Average Treatment Effects (CATE), $\hat{\tau}(X_i)$, and then calculates ATE by averaging them, as shown in Equation 7 (Wager and Athey 2018):

$$\tau_{ATE} = \frac{1}{n}\sum_{i=1}^{n} \hat{\tau}(X_i) \quad (7)$$

**Covariate Balance**

Validating the removal of selection bias requires assessing the distributional balance of covariates between the treated and control cohorts. This study utilizes the Standardized Mean Difference (SMD) (Austin 2009) as a scale-free diagnostic metric. For any specific covariate, the SMD is calculated as follows in Equation 8:

$$\text{SMD} = \frac{\mid \bar{X}_1 - \bar{X}_0 \mid}{\sqrt{\frac{S_1^2 + S_0^2}{2}}} \quad (8)$$

Here, $\bar{X}_1$ and $\bar{X}_0$ denote the sample means of the covariate in the treatment (2022 cohort) and control (2017 cohort) groups, respectively. $S_1^2$ and $S_0^2$ denote their corresponding sample variances. In observational causal inference, an absolute SMD threshold of 0.1 or below generally indicates that covariate balance has been successfully achieved.

**Heterogeneous Treatment Effects (HTEs) via Causal Forests**

While the ATE provides a population-level summary, it obscures distributional impact variances. Measuring these localized shifts requires estimating HTEs. The mathematical parameter of interest is the CATE, which defines the expected causal impact conditional on a specific covariate profile $x$ (Equation 9):

$$\tau(x) = E[Y_i(1) - Y_i(0) \mid X_i = x] \quad (9)$$

CATE is estimated using Causal Forests. Building upon the foundational architecture of Random Forests and Causal Trees, causal forests pivot away from minimizing standard predictive error. Instead, Athey and Imbens (2016) adapted the tree-splitting criterion to maximize heterogeneity in the treatment effect itself.

Wager and Athey (2018) expanded this concept into a forest ensemble. The algorithm recursively partitions the covariate space $X$ into terminal nodes. Within each causal tree, the localized treatment effect for any terminal leaf node $L$ containing a profile $x$ is estimated by contrasting the treated and control outcomes strictly within that leaf, as shown in Equation 10 (Athey and Imbens 2016):

$$\hat{\tau}(x) = \frac{1}{\mid \{i: W_i = 1, X_i \in L(x)\} \mid} \sum_{i: W_i = 1, X_i \in L(x)} Y_i - \frac{1}{\mid \{i: W_i = 0, X_i \in L(x)\} \mid} \sum_{i: W_i = 0, X_i \in L(x)} Y_i \quad (10)$$

where $L(x)$ denotes the terminal leaf containing the target covariate vector $x$, and $|\{i: W_i = w, X_i \in L(x)\}|$ represents the cardinality of units under treatment status $w$ residing within that specific leaf.

The leaf-specific treatment estimates across all trees in the ensemble are averaged by the causal forests (Wager and Athey 2018) (Equation 11):

$$\hat{\tau}(x) = \frac{1}{T}\sum_{t=1}^{T} \hat{\tau}_t(x) \tag{11}$$

Where $\hat{\tau}_t(x)$ is the treatment estimate from the terminal leaf containing x in tree t, and T is the total number of trees in the forest.

**Honest Estimation and Inference**

Traditional decision trees exhibit severe overfitting when assessing unobservable parameters due to adaptive splitting. This is the practice of using identical data to both construct the tree topology and compute the leaf parameters. Wager and Athey (2018) mitigate this via an architectural constraint termed the Honesty property.

During algorithm execution, the training data is randomly partitioned into a splitting subsample and an estimation subsample. The model utilizes the splitting subsample exclusively to determine feature thresholds and define the $L(x)$ geometry. It then calculates the actual causal parameters $\hat{\tau}(x)$ using only the orthogonal estimation subsample. Under regularity conditions, the honest estimation provides consistent estimates (Wager and Athey 2018).

**Conditional Average Treatment Effect (CATE)**

CATE profiles were calculated to estimate the variation in conditional treatment effect across observed covariate categories using the following equation (Equation 12):

$$\hat{\tau}_g = \frac{1}{n_g}\sum_{i:X_i=g} \hat{\tau}(X_i) \tag{12}$$

where $n_g$ represents the number of observations in category g. The estimated CATE profiles are accompanied by bootstrap confidence intervals computed by resampling the resulting CATE values within each subgroup.

**Feature importance**

The feature importance scores were extracted from the fitted causal forest model. These scores represent the relative contribution of the covariates in explaining heterogeneity in the estimated treatment effects (Wager and Athey 2018). A larger value of feature importance implies greater importance.

**Computational Tools**

All data ingestion and harmonization processes were executed via Python utilizing pandas and scikit-learn. The honest causal forest architecture was initialized and trained using the Econml library. Visual diagnostic outputs were synthesized with seaborn. Global random seeds were strictly declared to guarantee computational reproducibility across all stochastic tree-splitting and bootstrap resampling iterations.

## RESULTS

**Preliminary Data Analysis and Variable Independence**

Before applying the causal framework, we examined the foundational relationships between our study variables using a Spearman rank correlation matrix. Because our feature space consists primarily of ordinal categorical and binary variables, the Spearman method provides a more accurate nonparametric measure of monotonic relationships than the standard Pearson correlation.

As illustrated in Figure 1, the correlation values between any pair of independent covariates remain strictly below 0.5. The highest observed correlation in the feature space is between household vehicle count and household size, at 0.47, which is expected and falls below the commonly accepted multicollinearity threshold. It is worth noting that the correlation between the dependent outcome variables for car usage and walking is highly negative (-0.93). However, because the causal forest estimates treatment effects for each transport mode in entirely separate models, this high negative correlation in the outcome space does not confound the estimates.

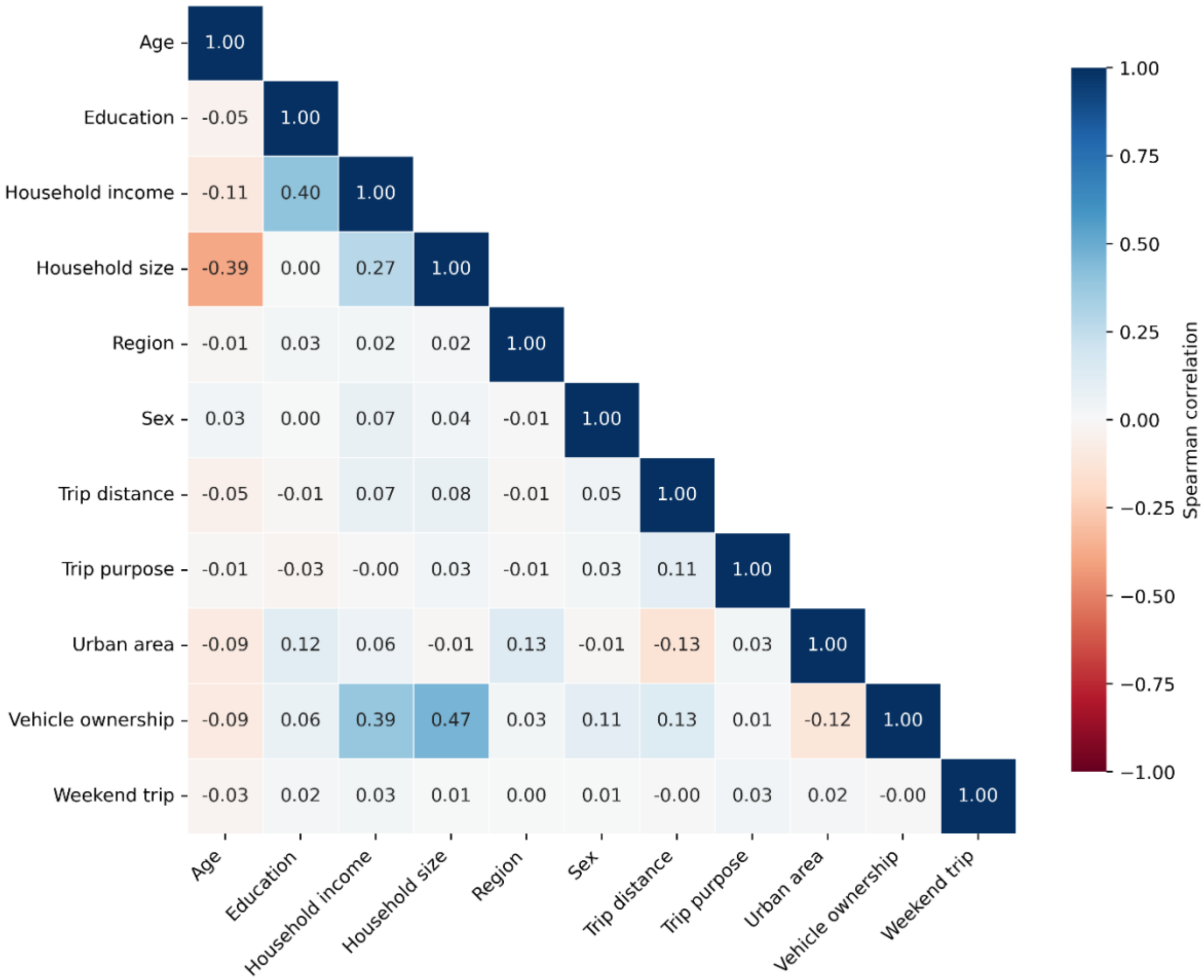

**Figure 1: Spearman correlation matrix of independent covariates and dependent mode choice variables.**

**Covariate Balance and the Standardized Mean Difference**
Validating the causal framework requires proving that our control group (2017) and treatment group (2022) are structurally comparable. We visualized the Standardized Mean Differences (SMD) across all independent variables using a Love plot (Figure 2).

Strict causal inference literature generally targets an absolute SMD threshold below 0.1 for perfect balance, though values up to 0.2 are frequently accepted in complex observational studies (Austin 2011). The Love plot reveals that almost all demographic and spatial variables fall comfortably within this acceptable threshold. We observe slight imbalances that exceed the strict 0.2 threshold in trip purpose and household income. Specifically, the 2017 baseline cohort had a slightly higher proportion of home-based trips and of higher-income households than the 2022 sample.

In traditional propensity score matching, this residual imbalance may require additional data trimming. This is not required for Causal Forest since tree-based ensemble methods natively partition the covariate space and adjust for localized imbalances during the splitting process; thus, the algorithm successfully neutralizes these slight distributional drifts without requiring the artificial destruction of valid survey data. This highlights the specific architectural advantage of the Causal Forest.

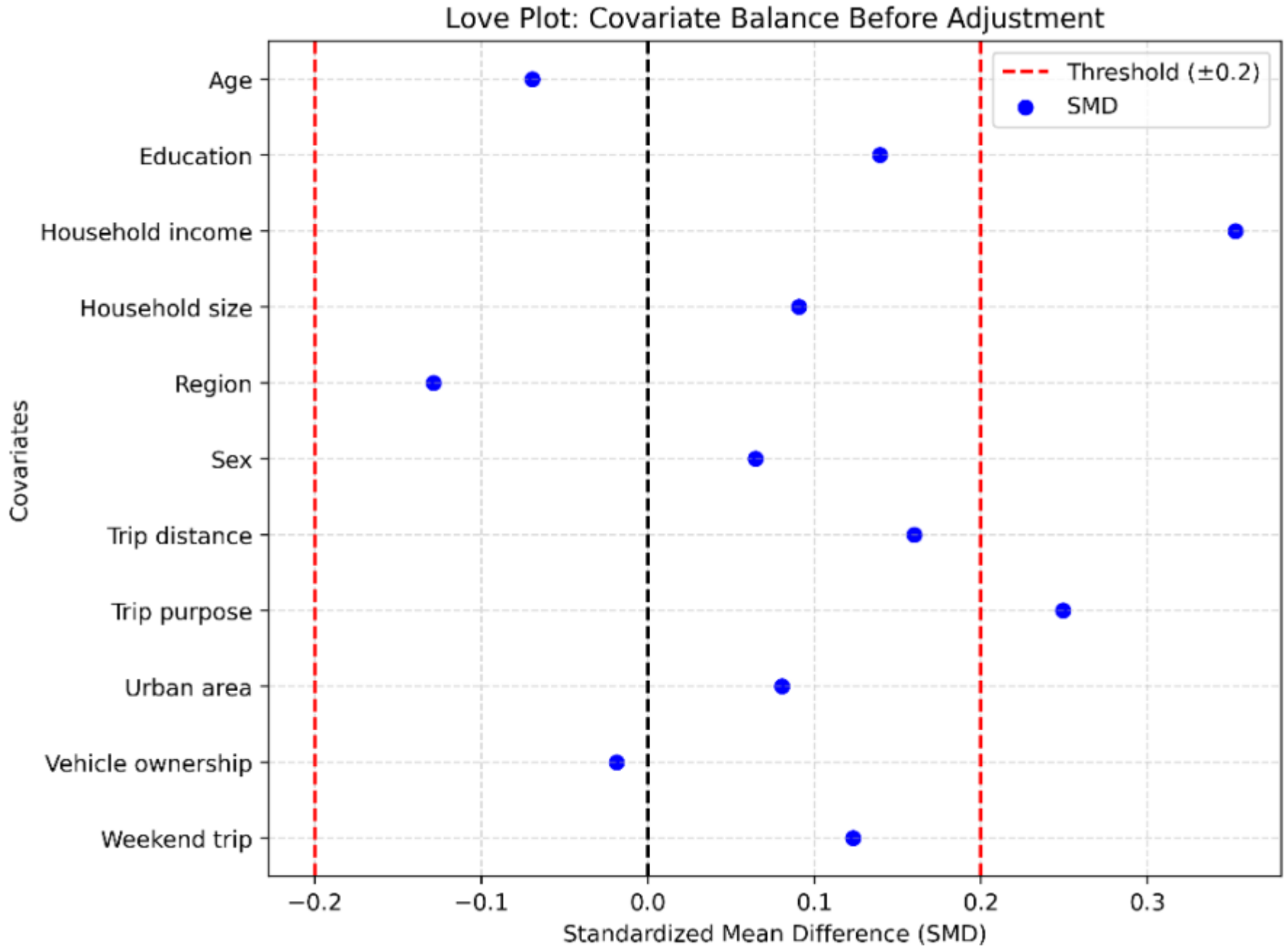

**Figure 2 Love plot demonstrating covariate balance via standardized mean differences between the 2017 and 2022 survey samples.**

**The Average Treatment Effect (ATE)**

Conditional on the observed covariates, the causal forest estimated that the transition from 2017 to 2022 increased the probability of car use by an average of 1.86 percentage points (pp), decreased the probability of public transit use by 0.38 pp, and decreased the probability of walking by 1.57 pp. It should be noted that these are average treatment effects; the distribution of estimated individual treatment effects across the sample (Figure 3) provides a more comprehensive view of treatment heterogeneity. The distribution indicates that, for at least a small proportion of the population, the transition from 2017 to 2022 was associated with reduced car use, increased public transit use, or increased walking, highlighting substantial heterogeneity in travel-behavior responses.

To quantify ATE precision, we used a non-parametric bootstrap resampling procedure with 1000 bootstrap samples. The bootstrapped standard errors were remarkably small (SE = 0.0001 across all three models), confirming that these ATE estimates are highly stable under the bootstrap sampling procedure:

I. Car: ATE: 0.0186, Bootstrap SE = 0.0001, Bootstrap 95% CI = [0.0183, 0.0188]
II. Public Transit: ATE: -0.0038, Bootstrap SE = 0.0001, Bootstrap 95% CI = [-0.0039, -0.0037]
III. Walking: ATE: -0.0157, Bootstrap SE = 0.0001, Bootstrap 95% CI = [-0.0160, -0.0155]

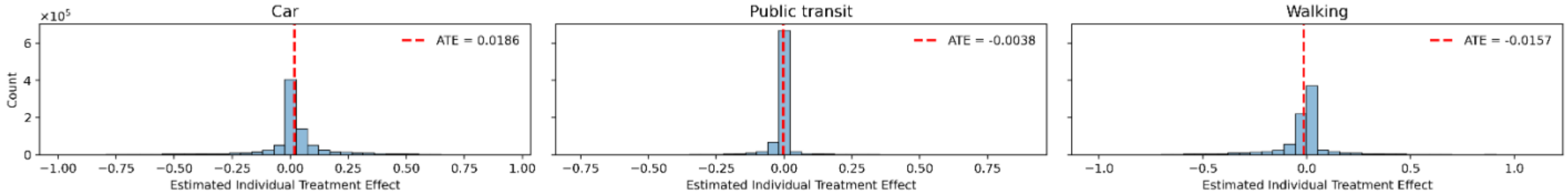

**Figure 3: Distribution of estimated average treatment effects across modes.**

**Robustness Across Estimation Methods**

To assess the robustness of the estimated ATE, the causal forest estimates were compared with those computed from an unadjusted difference-in-means estimator and a conventional multivariable linear regression. Across all three travel modes, the three approaches produced a consistent direction of the estimated effects and only slight variation in magnitude (Table 2).

**TABLE 2 Comparison of ATE Estimates Across Estimation Methods**

| Travel mode | Difference in Means (pp) | Adjusted Regression (pp) | Causal Forest (pp) |
|---|---|---|---|
| Car | 2.56 | 2.61 | **1.86** |
| Public Transit | -0.09 | -0.55 | **-0.38** |
| Walking | -2.47 | -2.06 | **-1.57** |

For both car use and walking, the causal forest estimated a slightly smaller ATE than both the unadjusted comparison and the adjusted regression. For public transit, all three approaches agreed on the direction and modest size of the effect, though the estimates varied more across methods, ranging from −0.09 to −0.55 pp, reflecting the small magnitude of the transit shift relative to the other modes.

**Heterogeneous Treatment Effects (HTE) Across Population Groups**

The global ATE masks severe behavioral disparities beneath the surface. To uncover the exact mechanics of the pandemic shock, we extracted the Conditional Average Treatment Effects (CATE) across specific sociodemographic categories. Figure 4 illustrates how the treatment effects varied across feature sub-levels.

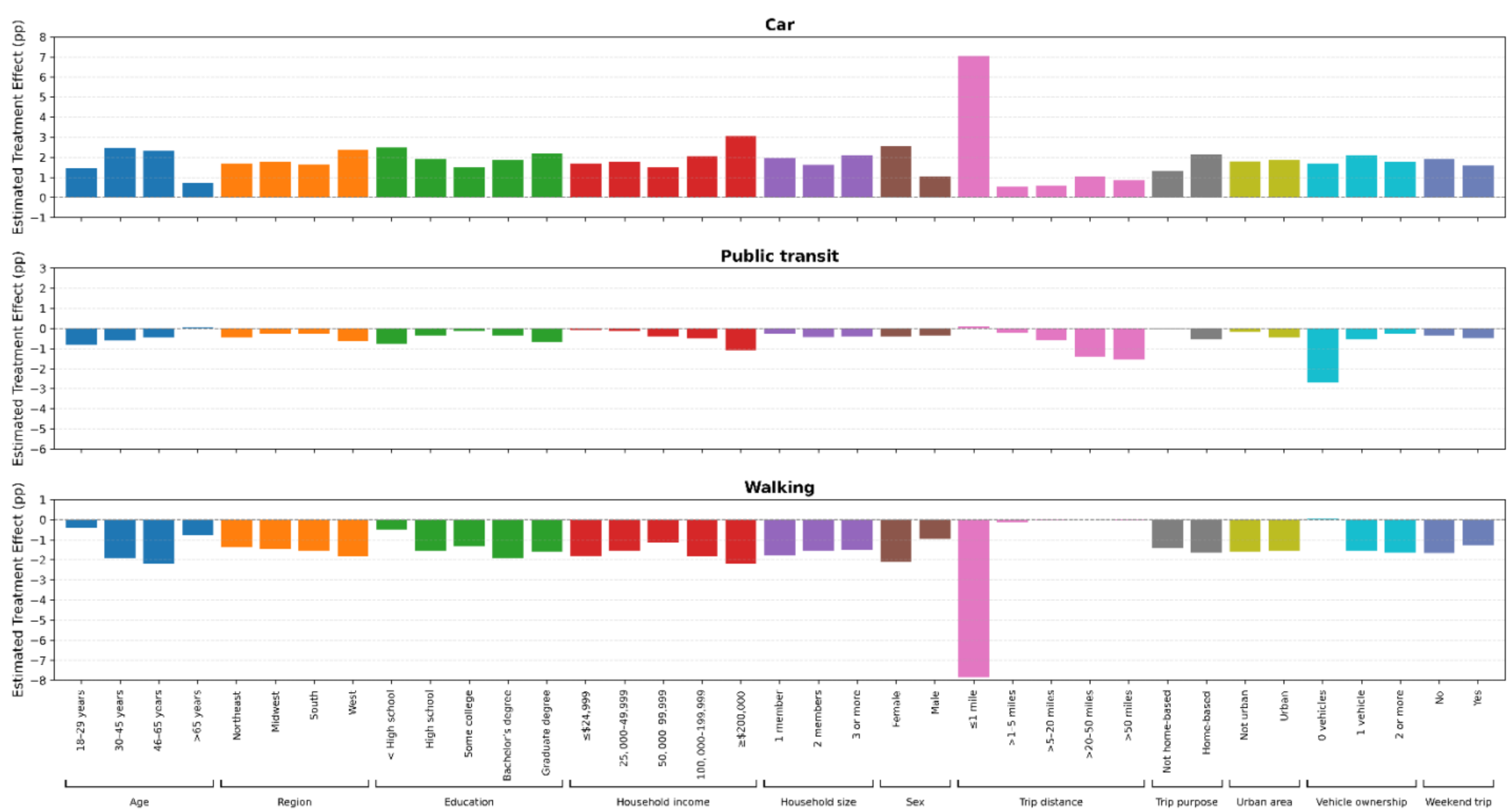

**Figure 4: Heterogeneous treatment effects isolated across distinct covariate subcategories.**

The algorithm exposes several critical demographic divergences, such as:

I. Trips of 1 mile or less witnessed the most drastic increase in car use (+7.06 pp) and decrease in walking (-7.85 pp). The estimated treatment effect on walking became increasingly negative with greater trip distance.

II. Households earning over $200,000 experienced a substantial increase in car usage (+3.06 pp) and the sharpest decrease in active walking (-2.21 pp).
III. Households with zero vehicles experienced a major drop in public transit usage (-2.71 pp) and a negligible shift toward walking (+0.06 pp).
IV. The data reveal that female travelers shifted away from walking (-2.01 pp) and towards cars (+2.55 pp) substantially more than males turned to cars (+1.04 pp).
V. The age groups 30- 45 years and 46–65 years drove the bulk of the car surge, +2.49 pp and +2.39 pp, respectively. The senior citizens (65+ years) exhibited deep behavioral inertia, with negligible change in mode use.
VI. Out of all Census divisions, the western U.S. experienced the largest increase in car use (+2.39 pp) and the largest decrease in public transit use (-0.63 pp) and walking (-1.82 pp).
VII. The change in mode use in urban versus rural U.S. was similar.
VIII. Work-related trips saw a larger increase in car use (+2.13 pp) and decrease in public transit use (-0.55 pp) than non-work-related trips (+1.33 pp and 0.05 pp, respectively).

**CATE Profiles Across Covariate Values**
To further understand treatment effect heterogeneity within each covariate category, CATE profiles were estimated. While the HTE analysis compared average treatment effects across population groups, CATE profiles illustrate how the estimated treatment effects (expressed in pp) vary across the observed levels of each covariate, thus providing insight into potential gradients and nonlinear relationships (Figure 5).

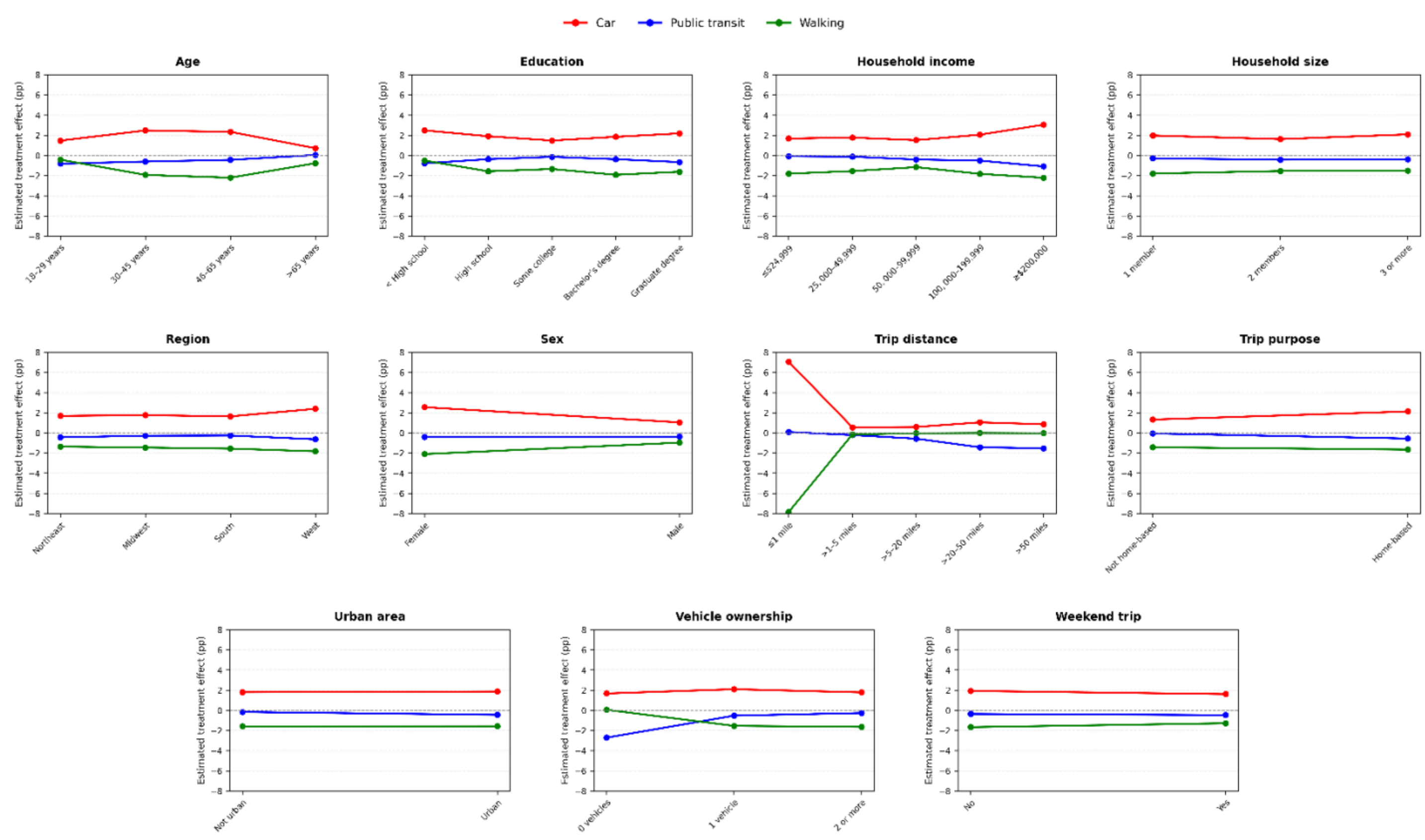

**Figure 5: CATE profiles across covariate values.**

**Variable Importance**
Variable importance was estimated using the causal forest model to identify covariates that contributed most strongly to treatment-level heterogeneity. These indicate only the relative contribution of each covariate in explaining the variation in estimated heterogeneous effects, rather than the direction and magnitude.

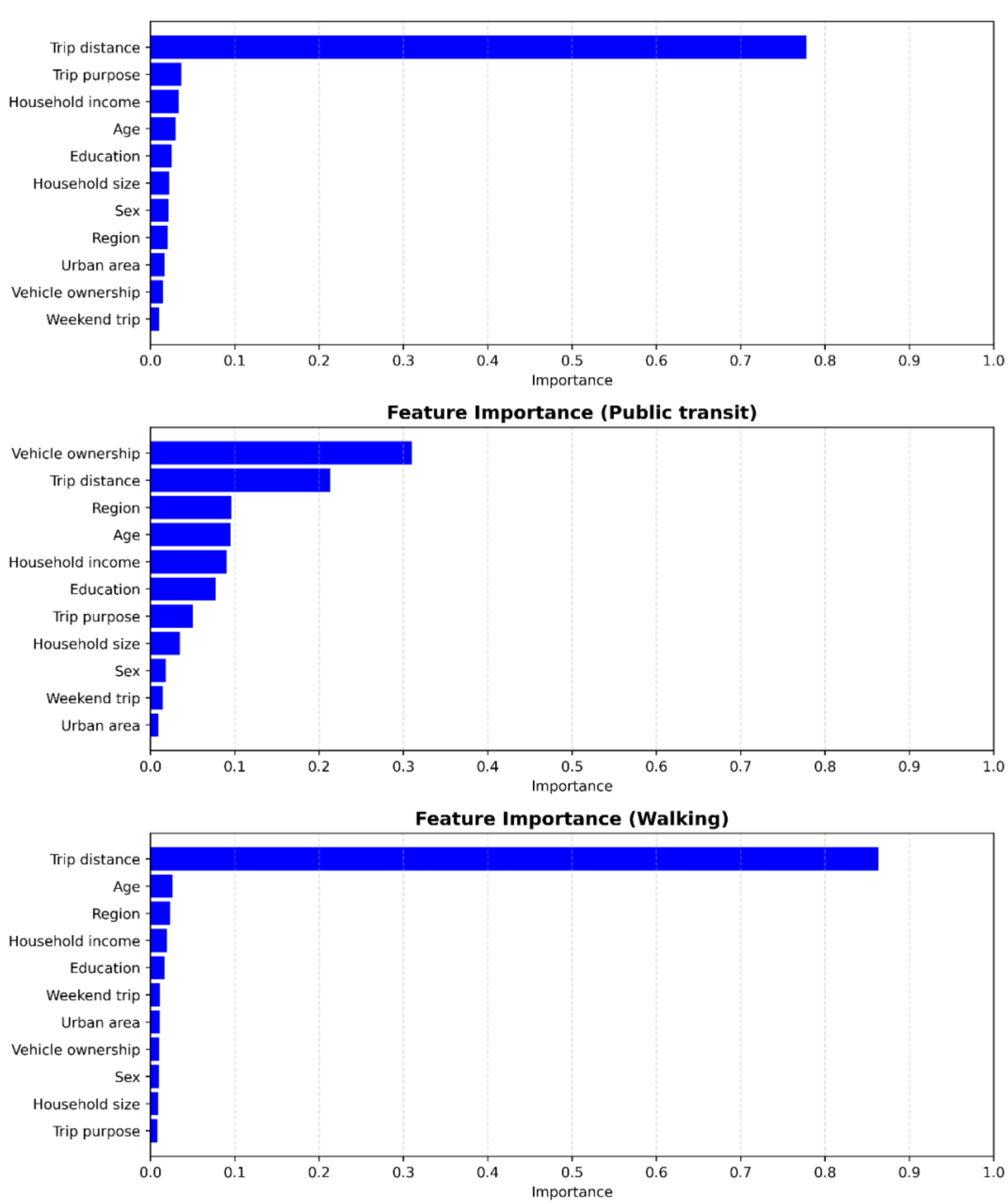

**Figure 6 Variable importance.**

As shown in Figure 6, trip distance was by far the most influential predictor of car use and walking, while household vehicles and trip distance were among the top contributors to public transit use.

**Exploratory Subgroup Analysis**

While ATEs report the overall effect of the pandemic on mode choice, they do not identify population group combinations with exceptionally different treatment effects. Thus, we conducted an exploratory subgroup analysis by evaluating all such combinations of up to three features and with more than 30 observations. The differences in estimated treatment effects between each feature combination and the remainder of the sample were assessed using Welch's two-sample t-tests, with statistical significance determined after Benjamini–Hochberg false discovery rate (FDR) correction ($\alpha = 0.05$). Table 3 presents feature combinations with the highest and lowest estimated treatment effects for each mode that are statistically significant after the FDR correction. It must be noted that this analysis is exploratory, not confirmatory.

**TABLE 3 Exploratory Subgroup Analysis**

| Target Mode | Intersecting Subgroup Profile | Size (n) | Group ATE | Rest of Sample ATE | Causal Difference | P value |
|---|---|---|---|---|---|---|
| Car (Highest increase) | West region, income \$25,000–\$49,999, trip ≤ 1 mile | 8442 | 18.05 | 1.68 | 16.37 | 0.00 |
| Car (Highest decrease) | No household vehicles, trip > 5 and ≤ 20 miles, non-home-based trip | 874 | -14.84 | 1.87 | -16.72 | 0.00 |
| Transit (Highest increase) | West region, no household vehicles, age 46–65 | 1573 | 8.58 | -0.40 | 8.98 | 0.00 |
| Transit (Highest decrease) | Income ≥ \$200,000, no household vehicles, trip > 1 and ≤ 5 miles | 126 | -21.18 | -0.38 | -20.80 | 0.00 |
| Walk (Highest increase) | Northeast region, no household vehicles, trip > 1 and ≤ 5 miles | 1640 | 21.67 | -1.62 | 23.30 | 0.00 |
| Walk (Highest decrease) | Income ≥ \$200,000, trip ≤ 1 mile, outside an urban area | 1369 | -17.49 | -1.55 | -15.94 | 0.00 |

## DISCUSSION

### Heterogeneity in Travel Behavior

The biggest advantage of using causal forests for mode choice analysis lies in their ability to capture heterogeneity in travelers' mode choice behavior. The ATEs can obscure substantial, localized behavioral disparities. In contrast, causal forests extend beyond ATEs, yielding a rich set of outputs, including HTE, CATE, and subgroup-level exploratory analyses. These insights are particularly valuable for policymakers, as they enable a more nuanced understanding of how interventions affect different population segments. For instance, our causal analysis estimated ATEs indicate a 1.86 pp increase in car-use mode share, alongside 0.38 and 1.57 pp decreases in public transit and walking mode shares, respectively. However, the HTE analysis reveals differences in mode choice shifts across trip characteristics and socio-demographics. In particular, short-distance trips (1 mile or less), households with incomes of $200,000 or more, and female travelers exhibit the largest increases in car-use mode share.

### Causal Forest Modeling of Travel Behavior

Observational travel behavior studies frequently fall into the correlation trap. They study the correlation or association between two variables but fail to identify causality (Chauhan 2023). By deploying a causal machine learning framework, this study goes beyond correlations to infer causality in shifts in travel behavior directly from observational data. Although this study demonstrated the use of causal forests to examine the causal effect of the COVID-19 pandemic on travel mode choice, the methodology can be applied to a wide range of travel behavior studies.

The causal forests offer several key advantages over traditional methods, for the following reasons:

i. Causal forests are designed explicitly for *causal inference* (Wager and Athey 2018). This alone is a major advancement over traditional travel mode choice studies, which are largely dominated by correlation-based methods (Chauhan, Riis, et al. 2024).
ii. They capture *heterogeneity* in causal treatment effects. This is particularly useful for travel behavior analysis since behavior may vary across individuals and groups.
iii. They offer very high *flexibility*. Causal forests are non-parametric and can capture complex non-linear relations (Athey and Wager 2019; Wager and Athey 2018). This makes travel behavior analysis data-driven and avoids strong assumptions on data structure.

iv. As demonstrated in this study, causal forests can be used for a wide *range of analysis*, including the estimation of ATE and its distribution, HTE across population groups, CATE profiles across covariate values, feature importance, and exploratory subgroup analysis. Each of these analyses offers a unique insight into shifts in travel behavior.

**Caveats to Causal Forest Result Interpretation**

The current study demonstrates how causal forests can be applied to mode choice analysis. There are important caveats to keep in mind when interpreting the results. These observations offer deeper insights into the findings and may also serve as important considerations for the future use of this methodology.

i. Overall, the estimated causal effects appear to be intuitively reasonable. Consistent with our findings, previous studies have also reported declines in public transit use and walking between the pre-pandemic period and 2022 (Clarridge et al. 2023; *Urban Transit Systems Labor Productivity* 2025). Similarly, researchers have noted the associational effect of demographic characteristics on mode choice (Magassy et al. 2024). However, it must be acknowledged that the effect of the pandemic on mode choice evolved over time, and thus, our findings should only be seen as a snapshot in time (Ciuffini et al. 2023; Javadinasr et al. 2022; Chauhan, Bhagat-Conway, et al. 2024).

ii. It is estimated that the average number of person trips declined substantially, from 36.1 trips per day in 2017 to only 23.9 trips per day in 2022 ("COVID-19 Related Transportation Statistics | Bureau of Transportation Statistics" 2022). Many workers switched to working from home, while some lost their jobs (Javadinasr et al. 2022). However, the current study focuses only on mode share rather than travel demand. Thus, the estimated change in mode use corresponds to the shift in mode share and not the absolute change in mode use.

iii. The estimated ATE, HTE, CATE, and other effects reported in this study are expressed in pp rather than in absolute terms. The practical significance of these estimates depends on the 2017 baseline values. For example, we estimate a treatment effect on car use of +1.86 pp and on walking of −1.57 pp. While these effects are similar in magnitude, baseline car use in 2017 is overwhelmingly higher than walking. Consequently, a change of similar magnitude can have a disproportionately larger impact when the baseline value is higher.

iv. Some covariates used in this study, including the number of household vehicles and household income, may themselves have been affected by the pandemic. The causal forest methodology and the cross-sectional data used in this study cannot capture such causal effects. A more sophisticated methodology, or longitudinal data following the same individuals over time (Chauhan, Bhagat-Conway, et al. 2024), would be required to take this additional step. The current study does not account for these potential effects.

v. We found that the ATEs estimated using causal forests differed only modestly in magnitude from those obtained via differences in means and adjusted regression, while the directions of the effects were consistent across all three methods. Nevertheless, the primary advantage of causal forests lies in their ability to estimate heterogeneous treatment effects, which are not captured by aggregate estimates.

**Assumptions of the Study**

As with other observational causal inference methods, machine learning approaches, and conventional statistical models, the treatment effects estimated by causal forests should be understood as model-based estimates contingent on the analysis's underlying assumptions. That said, because causal forests are data-driven and non-parametric, they rely on fewer assumptions. Below are the main assumptions of our methodology:

i. *Conditional exchangeability*: It assumes that there are no unmeasured variables (confounders), beyond the covariates already included, that affect both the survey year and travel mode choice. While the study included several key socio-demographic and trip-related covariates, this assumption may still be the most stringent. Factors such as remote working, online learning, online shopping, urban exodus, attitudes, and risk perception may be relevant but were not included (Chauhan, Capasso Da Silva, et al. 2021; Salon, Conway, Silva, et al. 2021; Capasso da Silva et al. 2021; Mirtich et al. 2021). Future studies must pay extra attention to the extent of covariates included.

ii. *Positivity or overlap*: It assumes that all combinations of characteristics must be represented in both the treatment as well as the control group. With our final dataset of over 800K trips, this assumption is likely satisfied. This may, however, be a concern for smaller datasets.

iii. SUTVA (Stable Unit Treatment Value Assumption): It assumes that one individual's treatment does not affect another individual's outcome, and that the treatment is administered consistently. The validity of the assumption may depend primarily on the sampling method and the treatment under study.

## CONCLUSION

Causal models have under-explored potential in transport planning (Chauhan 2023). Recent research has demonstrated the value of causal discovery and inference in mode choice analysis (Chauhan, Riis, et al. 2024; Chauhan et al. 2026, 2025), yet heterogeneity in causal analysis remains unexamined. This gap is critical because heterogeneity is central to understanding travel behavior and to informing travel-related policies, as it reveals variability in behavior across individuals and population groups.

In this study, we demonstrate the use of causal forests to estimate heterogeneity in shifts in travel behavior caused by the COVID-19 pandemic, using NHTS travel survey data collected both before and during the pandemic. Our findings reveal substantial variation in mode choice shifts across socio-demographic and trip characteristics. Causal forests, which are causality-based, data-driven, and nonparametric, emerge as a promising modeling approach for studying heterogeneity in shifts in travel mode choice. They provide not only ATE estimates of the pandemic's impact on mode choice but also HTE, CATE, feature importance, and exploratory subgroup analysis, enabling an in-depth understanding of shifts in travel behavior.

## ACKNOWLEDGMENTS

None

## AUTHOR CONTRIBUTIONS

Study conception and design: RSC. Data collection and processing: LLG, RSC. Analysis and interpretation of results: RSC, MG, LLG. Draft manuscript preparation: RSC, MG, LLG. All authors reviewed the results and approved the final version of the manuscript.

## DATA AVAILABILITY STATEMENT

The data used in this study are publicly available. The 2017 and 2022 National Household Travel Survey datasets can be accessed at https://nhts.ornl.gov. No proprietary or restricted data were used.

## DECLARATION OF GENERATIVE AI USE

During the preparation of this manuscript, the authors used ChatGPT (OpenAI) to assist with code debugging and language editing. All scientific content, analyses, interpretations, and conclusions were developed and verified by the authors, who take full responsibility for the manuscript.

## DECLARATION OF CONFLICTING INTERESTS

The authors declare no potential conflicts of interest with respect to the research, authorship, or publication of this article.

## FUNDING

The authors disclosed no financial support for the research, authorship, and/or publication of this article.